\documentclass[5p,10pt]{elsarticle}

\usepackage{subfigure}
\usepackage{multirow}
\usepackage{amsmath}
\usepackage{amssymb}
\usepackage{setspace}
\usepackage{mdwlist}
\usepackage{epsfig}

\def\be{\begin{equation}}
\def\ee{\end{equation}}

\begin{document}

\begin{frontmatter}

\author[1]{L. A. Harland-Lang}
\author[1,2]{V. A. Khoze}
\author[2]{M. G. Ryskin}
\author[3]{W. J. Stirling}
\address[1]{Department of Physics and Institute for Particle Physics Phenomenology, University of Durham, DH1 3LE, UK}
\address[2]{Petersburg Nuclear Physics Institute, NRC Kurchatov Institute, Gatchina, St. Petersburg, 188300, Russia}
\address[3]{Cavendish Laboratory, University of Cambridge, J.J. Thomson Avenue, Cambridge, CB3 0HE, UK}

\title{\begin{flushright}{\rm \small
IPPP/13/21, DCPT/13/42, Cavendish-HEP-13/04}\\\hbox{$\null$\break}
\end{flushright}
Probing the perturbative dynamics of exclusive meson pair production}

\begin{abstract}
 We present the results of a recent novel application of the `hard exclusive' perturbative formalism to the process \linebreak[4]$gg \to M\overline{M}$ at large angles, where $M$ ($\overline{M}$) is a light meson (anti--meson). As well as discussing the important theoretical features of the relevant leading--order $gg \to q\overline{q}(gg) q\overline{q}(gg)$ 6--parton amplitudes, we also comment on their phenomenological implications. In particular, we consider the central exclusive production of meson pairs at comparatively large transverse momentum $k_\perp$, which is expected to proceed via this mechanism. We show that this leads to various non--trivial predictions for a range of exclusive processes, and that the cross sections for the $\eta'$ and $\eta$ mesons display significant sensitivity to any valence $gg$ component of the meson wavefunctions.
\end{abstract}

\end{frontmatter}

\section{Introduction}

Central exclusive production (CEP) processes of the type
\begin{equation}\label{exc}
pp({\bar p}) \to p+X+p({\bar p})\;,
\end{equation}
can significantly extend the physics programme at hadron colliders. Here $X$ represents a system of invariant mass $M_X$, and the `$+$' signs denote the presence of large rapidity gaps. Such reactions represent an experimentally very clean signal and provide a very promising way to investigate both QCD dynamics and new physics in hadron collisions, and consequently they have been widely discussed in the literature: we refer the reader to~\cite{Khoze:2001xm,Martin:2009ku,Albrow:2010yb,HarlandLang:2013jf}  for reviews and further references. 

One important feature of such reactions, which we will make use of later on, is the dynamical `$J_z^{PC}=0^{++}$ selection rule': that is, the produced state must have even $C$ parity, while the production of states with non--$J_z^P=0^+$ quantum numbers (where $J_z$ is the projection of the object angular momentum on the beam axis) is strongly suppressed~\cite{Khoze:2000jm,Kaidalov:2003fw}. This suppression is predicted in the Durham pQCD--based model to be at the level of two orders of 
magnitude.

\begin{figure}[h]
\begin{center}
\includegraphics[scale=1.0]{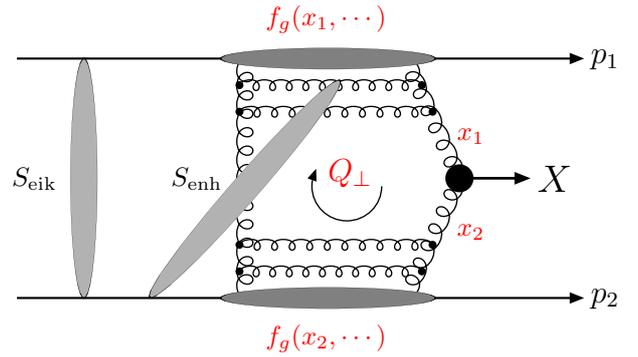}
\caption{The perturbative mechanism for the exclusive process $pp \to p\,+\, X \, +\, p$.}
\label{fig:pCp}
\end{center}
\end{figure} 

A particularly interesting example, which has been the topic of recent investigations in~\cite{HarlandLang:2011qd,HarlandLang:2013}, is the production of light meson pairs ($X=\pi\pi, KK, \rho\rho, \eta(')\eta(')$...) at sufficiently high transverse momentum $k_\perp$ that a perturbative approach, applying the Durham pQCD--based model of CEP (see e.g.~\cite{HarlandLang:2013jf} and references therein) and the `hard exclusive' formalism (see e.g.~\cite{Brodsky:1981rp,Benayoun:1989ng}), to evaluate the meson production subprocess, can be taken. 

There are two principle, related, reasons to look at such reactions. First, the parton--level helicity amplitudes relevant to the CEP of meson pairs, $gg \to M\overline{M}$ (where $M,\overline{M}$ is a meson, anti--meson) exhibit some remarkable theoretical features. Such `exclusive' $gg \to q\overline{q}(gg) q\overline{q}(gg)$ 6--parton amplitudes with fixed helicities of the incoming gluons are not relevant in a typical high--multiplicity inclusive process, and consequently they have not been studied in this context before. Second, they are of phenomenological interest, being experimentally realistic observables at hadron colliders from the Tevatron and RHIC to the LHC, with the wide variety of meson states available offering various channels with which to probe the non--trivial theory predictions for these different processes. In this paper we will discuss these two aspects in turn, emphasising the most important features and implications of our approach, without going into the detailed treatment of~\cite{HarlandLang:2011qd,HarlandLang:2013}.

\section{Properties of the $gg \to M \overline{M}$ amplitudes}

\begin{figure}
\begin{center}
\includegraphics[scale=1.2]{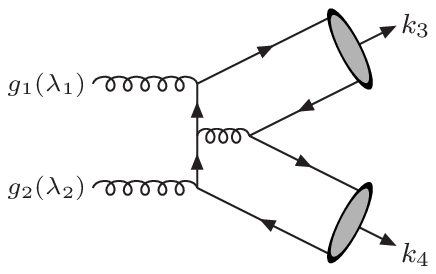}
\caption{Representative Feynman diagram for the $gg \to q\overline{q}q\overline{q}(gg)$ process.}\label{feyn1}
\end{center}
\end{figure}

The leading order amplitudes are calculated using an extension of the `hard exclusive' formalism, see~\cite{Brodsky:1981rp,Benayoun:1989ng}. The basic idea is that the hadron--level amplitude can be written as a convolution of a (perturbatively calculable) parton--level amplitude, $T$, and a `distribution amplitude' $\phi$, which contains all the (non--perturbative) information about the binding of the partons in the meson. The $gg \to M\overline{M}$ amplitude can be written as
\begin{equation}\label{amp}
\mathcal{M}_{\lambda\lambda'}(\hat{s},\theta)=\int_{0}^{1} \,{\rm d}x \,{\rm d}y\, \phi_M(x)\phi_{\overline{M}}(y)\, T_{\lambda\lambda'}(x,y;\hat{s},\theta)\;,
\end{equation}
where $\sqrt{\hat{s}}$ is the $M\overline{M}$ invariant mass, $x,y$ are the meson momentum fractions carried by the partons and $\theta$ is the scattering angle in the $gg$ cms frame. $T_{\lambda\lambda'}$ is the hard scattering amplitude for the parton--level process \linebreak[4]$gg\to q\overline{q}(gg)\,q\overline{q}(gg)$, where each $q\overline{q}$ or $gg$ pair is collinear with the meson momentum\footnote{For a meson produced with large momentum, $|\vec{k}|$, we can to good approximation neglect the transverse component of the parton momentum, $\vec{q}$, with respect to $\vec{k}$.} and has the appropriate colour, spin, and flavour content projected out to form the parent meson. $\lambda$, $\lambda'$ are the gluon helicities: for our considerations there are two independent helicity configurations, $(\pm\pm$) and $(\pm \mp)$, which correspond to the incoming gluons being in a $J_z=0$ and $|J_z|=2$ state, respectively, along the incoming $gg$ direction. A representative diagram for purely $q\overline{q}$ valence components is shown in Fig.~\ref{feyn1}. Provided the meson $k_\perp$ is large enough, all intermediate quark and gluon propagators will be far off--shell and the amplitude can be calculated using the standard tools of pQCD.

\begin{figure}
\begin{center}
\includegraphics[scale=0.75]{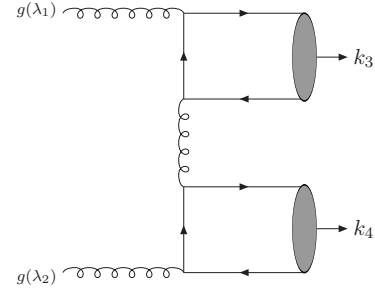}
\caption{Representative `ladder type' diagram, which contributes for flavour--singlet mesons.}\label{feyn2}
\end{center}
\end{figure}

Originally, in~\cite{Brodsky:1981rp,Benayoun:1989ng}, the simpler case of initial--state photons, $\gamma\gamma \to M \overline{M}$, was considered. In~\cite{HarlandLang:2011qd,HarlandLang:2013} we extended these results to the two--gluon case. Taking the simplest case of scalar flavour--non--singlet ($\pi\pi, KK$...) production, this proceeds via the type of diagram shown in Fig.~\ref{feyn1}. We will present the results first, before commenting on them at the end of the section. There are 31 Feynman diagrams which contribute to the leading--order amplitude, and after an explicit calculation we find
\begin{align}\label{T++}
T^{qq}_{++}=T^{qq}_{--}&=0\;,\\ \label{T+-}
T^{qq}_{+-}=T^{qq}_{-+}&=\frac{\delta^{\rm AB}}{N_C}\frac{64\pi^2\alpha_S^2}{\hat{s}xy(1-x)(1-y)}\frac{a-b^2}{a^2-b^2\cos^2{\theta}}\\ \nonumber
&\cdot \frac{N_C}{2}\bigg(\cos^2{\theta}-\frac{2 C_F}{N_C}a\bigg)\;,
\end{align}
where `$qq$' indicates that the final--state partons are $q\overline{q}$ pairs, `$A,B$' are the gluon colour indices and
\begin{align}\label{a}
a&=(1-x)(1-y)+xy\; ,\\ \label{b}
b&=(1-x)(1-y)-xy\; .
\end{align}
As well as the configuration shown in Fig.~\ref{feyn1}, the outgoing $q\overline{q}$ pairs can also combine in a second way, with the $q\overline{q}$ pair forming each meson being connected by a quark line. A representative such `ladder--type' diagram is shown in Fig.~\ref{feyn2}; as each meson couples individually to two (isosinglet) gluons, only flavour--singlet states can be produced via such a diagram (for e.g. a $\pi^\pm$ state, this is clear, while for a $\pi^0$ the $u\overline{u}$ and $d\overline{d}$ components of the flavour Fock state interfere destructively), while for the case of initial--state photons it is forbidden in all cases by colour conservation. There are 8 Feynman diagrams which contribute to the amplitude, and we find that these give
\begin{align}\label{lad0}
T_{++}^{S,qq}=T_{--}^{S,qq}&=\frac{\delta^{ab}}{N_C}\frac{64\pi^2 \alpha_S^2}{\hat{s}xy(1-x)(1-y)}\frac{(1+\cos^2 \theta)}{(1-\cos^2 \theta)^2}\;,\\ \label{lad2}
T_{+-}^{S,qq}=T_{-+}^{S,qq}&=\frac{\delta^{ab}}{N_C}\frac{64\pi^2 \alpha_S^2}{\hat{s}xy(1-x)(1-y)}\frac{(1+3\cos^2 \theta)}{2(1-\cos^2 \theta)^2}\;,
\end{align}
where the label `$S,qq$' is used to distinguish these amplitudes, which only contribute for flavour--singlet mesons, from (\ref{T++}, \ref{T+-}), which contribute for both flavour--singlet and non--singlet states.

As well as having valence $q\overline{q}$ components, it is well known that the dominantly flavour--singlet $\eta'$ (and also, through mixing, $\eta$) mesons should have a valence $gg$ component, which also carries flavour--singlet quantum numbers. While different determinations of the $\eta$--$\eta'$  mixing parameters are generally consistent, the size of the gluonic content of the $\eta'$ ($\eta$) mesons has remained uncertain for many years, with different fits giving conflicting results, see for example~\cite{Thomas:2007uy,Kroll:2012hs} (as well as~\cite{HarlandLang:2013} and references therein) for a discussion of the theoretical uncertainties and~\cite{DiDonato:2011kr} for a review of the experimental situation. With this in mind, we can extend the above calculation to include the case that one or both outgoing $q\overline{q}$ pairs forming the meson states are replaced by $gg$ pairs in a pseudoscalar state. Such diagrams will contribute to the $gg \to \eta(')\eta(')$ processes in the presence of any non--zero $gg$ valence component. We find
\begin{align}\label{tgq0}
 T^{gq}_{++}&=T^{gq}_{--}=2\,\sqrt{\frac{N_C^3}{N_C^2-1}}(2x-1)\cdot\,T_{++}^{S,qq}\;,\\ \label{tgg0}
T^{gg}_{++}&=T^{gg}_{--}=4\,\frac{N_C^3}{N_C^2-1}(2x-1)(2y-1)\cdot\,T_{++}^{S,qq}\;,
\end{align}
with 76 Feynman diagrams contributing to (\ref{tgq0}) and 130 to (\ref{tgg0}). We can find no simple form in the case of the \linebreak[4]$|J_z|=2$ ($\pm,\mp$) amplitudes, but the angular dependences are found numerically to be similar.

We can see that these amplitudes have some remarkable features, which we summarize below:

\begin{enumerate}
 \item The quark amplitude corresponding to the case that the $q\overline{q}$ pair forming each meson are not connected by a quark line, as in Fig.~\ref{feyn1}, completely vanishes for the case that the incoming gluons are in a $J_z=0$ state, see (\ref{T++}). This non--trivial result follows from an overall cancellation between the many different non--zero contributing Feynman amplitudes, and is a direct generalization of the case of the equivalent $\gamma\gamma \to M \overline{M}$ amplitudes, which vanish for neutral mesons, when the photons are in a $J_z=0$ state~\cite{Brodsky:1981rp}.
 \item The quark amplitude (\ref{lad0}) corresponding to the case that the $q\overline{q}$ pair forming each meson {\it is} connected by a quark line, as in Fig.~\ref{feyn2}, and which only contributes for flavour--singlet mesons, does not vanish for the case of $J_z=0$ incoming gluons. Thus we observe, from this simple fact, a strikingly different behaviour between the cases of flavour singlet and non--singlet meson pair production. As we will see, this has important phenomenological consequences for meson pair CEP.
 \item If we replace the $q\overline{q}$ pairs forming the flavour--singlet meson states by gluons, then again the amplitudes (\ref{tgq0}, \ref{tgg0}) for $J_z=0$ incoming gluons do not vanish. Moreover, they are in fact identical to the amplitude (\ref{lad0}) for a purely valence quark component in the mesons, up to overall colour and normalization factors. That is, they are predicted to have the same angular dependence in the incoming $gg$ rest--frame. We emphasize that in these cases, all diagrams allowed by colour conservation contribute to the total amplitude, and not just diagrams of this ladder type shown in Fig.~\ref{feyn2}: for each process (`$S,qq$', `$qg$' and `$gg$') the final amplitude receives contributions from a set of diagrams which are completely distinct and apparently unrelated between the three cases. That these final results should be so remarkably similar in form is therefore very surprising.
 \item The amplitude (\ref{T+-}), corresponding to the diagram of the type shown in Fig.~\ref{feyn1}, for $|J_z|=2$ incoming gluons, vanishes for a particular value of $\cos^2\theta$. This behaviour, which at first sight may appear quite unusual, is in fact not completely unexpected: the vanishing of a Born amplitude for the radiation of massless gauge bosons, for a certain configuration of the final state particles, is a known effect, usually labeled a `radiation zero', see for instance~\cite{Heyssler:1997ng,Brown:1982xx}. The position of the zero is determined by an interplay of both the internal (in the present case, colour) and space-time (the particle $4$-momenta) variables, as can be seen in (\ref{T+-}), where the position of the zero depends on the choice of meson wavefunction, $\phi(x)$, through the variables $a$ and $b$, as well as on the QCD colour factors. 
 \item The parton--level amplitudes corresponding to the $gg \to M\overline{M}$ process can also be understood by a novel and instructive application of the MHV formalism (see for example~\cite{Mangano:1990by} for a review and~\cite{Georgiou:2004wu} for a more recent reference). In particular, these amplitudes for $J_z=0$ gluons are MHV, so by taking the known results for the general 6--parton MHV amplitudes and substituting in the kinematic and colour information relevant to the specific two--meson production process, this allows the previous results to be derived in a relatively simple way, without resorting to a Feynman diagram calculation. In this way, the vanishing of the valence quark amplitude (\ref{T++}) for the case of $J_z=0$ incoming gluons follows from a few lines of calculation. The identical form of the three amplitudes (\ref{lad0}, \ref{tgq0}, \ref{tgg0}) may also be explained in the MHV framework by the fact that the same external parton orderings in the MHV amplitudes contribute in all three cases. The $|J_z|=2$ amplitude (\ref{T+-}), for example, may also be derived using the `MHV rules', see~\cite{Georgiou:2004wu} for references. We refer the reader to~\cite{HarlandLang:2011qd,HarlandLang:2013} for more technical details.
 \item Although we do not show them explicitly here, the valence--quark helicity amplitudes for  vector mesons ($\rho\rho$, $\omega\omega$, $\phi\phi$) have also been calculated in~\cite{HarlandLang:2011qd}. As in (\ref{T++}), for the case that the $q\overline{q}$ pair forming each meson are not connected by a quark line, the $J_z=0$ amplitude vanishes. Transversely polarized $\lambda=\pm 1$ vector mesons cannot be produced within this formalism via the diagram of the type shown in Fig.~\ref{feyn2}, due to helicity conservation along the quark line, while an explicit calculation shows that the amplitude vanishes for the case of longitudinal polarizations.
\end{enumerate}

\section{Phenomenological implications}

We recall that the CEP of meson pairs at sufficiently high $k_\perp$ can be modelled using the Durham pQCD--based model for the exclusive production of a generic object $X$, produced via the $gg \to X$ subprocess, see e.g.~\cite{Khoze:2001xm,Khoze:2000jm,Kaidalov:2003fw} for general references, as well as~\cite{HarlandLang:2011qd,HarlandLang:2010ep} for a detailed account of the model that we use here. The case we are interested in is $X=M\overline{M}$, for which the amplitudes calculated in the previous section can be used, up to corrections due to the fact that the incoming gluons are not completely on--shell: for the reasonably high invariant masses, $M_X$, we are considering here, these will in general be small~\cite{HarlandLang:2011qd}. The soft survival factor is calculated using the model described and referenced to in~\cite{HarlandLang:2010ep}, with the size of the overall suppression factor given by $\langle S^2 \rangle \approx 3$--$10\%$, depending on the process considered and c.m.s. energy. We list below the most important phenomenological implications of this in the context of meson pair CEP, before briefly discussing the relevance of a separate `non--perturbative' picture in the lower $k_\perp$ region, and the implications this has for the existing exclusive data.

\begin{enumerate}
 \item We have seen from (\ref{T++}) that the quark amplitudes corresponding to the case that the $q\overline{q}$ pair forming each meson are not connected by a quark line, as in Fig.~\ref{feyn1}, completely vanishes for the case that the incoming gluons are in a $J_z=0$ state. At leading order, the production of flavour non--singlet meson pairs, which must proceed via these diagrams, can therefore only occur when the incoming gluons are in a $|J_z|=2$ state. Recalling the `$J_z=0$' selection rule discussed in the introduction, which strongly disfavours such a configuration, this gives the highly non--trivial prediction that the CEP of flavour non--singlet meson pairs ($\pi\pi, KK$...) in the perturbative regime, will be strongly suppressed\footnote{The fusing gluons in CEP can to good approximation be treated as being on--shell, in which case the $gg$ and beam axis coincide, and so the $J_z$ of the gluons in the $gg$ rest--frame must be equal to the projection of the centrally produced object's angular momentum on the beam axis.}. We note that e.g. the $\pi^+\pi^-$ cross section is simply predicted to be twice the neutral $\pi^0\pi^0$ cross section, due to the non--identity of the final state particles. 
 \item The above result has an important implication for the case of $\gamma\gamma$ CEP, proceeding via the quark box $gg \to \gamma\gamma$ diagram, see for example~\cite{HarlandLang:2012qz}. In general the background from exclusive $\pi^0\pi^0 \to 4 \gamma$ production, when one photon from each $\pi^0$ decay is undetected or the two photons merge, may be important: we would naively expect the cross section for such a purely QCD process to be larger than in the direct $\gamma\gamma$ case. However, if we consider for example the cross section at $\sqrt{s}=1.96$ TeV, for the pions/direct photons restricted to have $E_\perp>2.5$ GeV and $|\eta|<1$, then we predict $\sigma(\pi^0\pi^0)/\sigma(\gamma\gamma) \approx 1\%$. Without the suppression due to the $J_z=0$ vanishing of (\ref{T++}), we would instead expect the cross sections to be roughly equal. In~\cite{Aaltonen:2011hi} the observation of 43 $\gamma\gamma$ events at $\sqrt{s}=1.96$ TeV with $|\eta(\gamma)|<1.0$ and  $E_T(\gamma)>2.5$ GeV, with no other particles detected in $-7.4<\eta<7.4$ was reported by CDF, and the contamination caused by $\pi^0\pi^0$ production was determined experimentally to be  very small, and consistent with zero (finding $N(\pi^0\pi^0)/N(\gamma\gamma)<0.35$, at 95\% C.L.), which provides good support for our approach. An observation of $\pi^0\pi^0$ CEP which may hopefully come with the increased statistics that a further analysis of the existing CDF data can bring~\cite{albrow,MAdif12}, would certainly represent an interesting further test of the theoretical formalism.
 
 \begin{figure}[h]
\begin{center}
\includegraphics[scale=0.7]{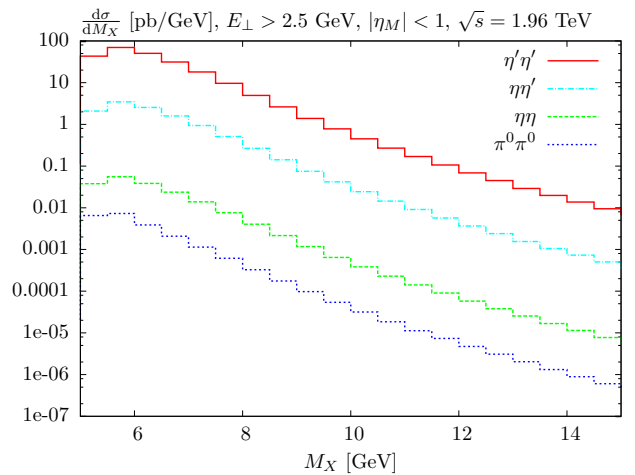}
\caption{${\rm d}\sigma/{\rm d}M_X$ for meson transverse energy $E_\perp>2.5$ GeV and pseudorapidity $|\eta_M|<1$ at $\sqrt{s}=1.96$ TeV for the CEP of meson pairs, calculated within the perturbative framework, with no $gg$ valence component included for the $\eta(')$ mesons. MSTW08LO PDFs are used, and the $q\overline{q}$ distribution amplitude at starting scale $\mu_0=1$ GeV is given by the CZ form (see~\cite{Benayoun:1989ng}).}\label{vsm}
\end{center}
\end{figure}
 
 \item We have seen that the quark amplitudes corresponding to the case that the $q\overline{q}$ pair forming each meson {\it are} connected by a quark line, as in Fig.~\ref{feyn2}, do not vanish for $J_z=0$ incoming gluons. We recall that these type of diagrams can only contribute for flavour--singlet mesons. From this simple observation, we then have the highly non--trivial prediction that the CEP of flavour--singlet meson pairs, in particular $\eta'\eta'$, will not be suppressed by the $J_z=0$ selection rule, and so it is predicted to be strongly enhanced relative to e.g. $\pi\pi$ CEP. In the case of $\eta\eta'$ and $\eta\eta$ production, we also expect some enhancement, although the level of this is dependent on the specific $\eta$--$\eta'$ mixing parameters that are taken. An observation of these $\eta(')\eta(')$ CEP cross sections, or of the ratio of $\sigma(\eta(')\eta('))$ to $\sigma(\pi\pi)$ or $\sigma(\gamma\gamma)$, for which various uncertainties (due to PDF choice, soft survival factors etc) cancel out, would therefore represent another interesting test of this formalism: we show representative predictions for such ratios in Table~\ref{etarats1}, with $a_2^G(\mu_0^2)=0$ corresponding to the case of purely $q\overline{q}$ valence $\eta(')$ components considered so far. We emphasize in particular that the $\eta'\eta'$ CEP cross section is predicted to be very large, and that the branching ratios for the  $\rho^0\gamma$ and $\eta \pi^+\pi^-$ decays of the $\eta'$ are sizeable, and may be viable channels. In Fig.~\ref{vsm} we plot the predicted $M_X$ distribution for a range of meson states, and the relative enhancement and suppression of the $\eta'\eta'$ and $\pi\pi$ cross sections, respectively, is clear. In principle $\eta(')\eta(')$ CEP followed by $\eta(')\to \gamma\gamma$ could therefore represent a further background to exclusive $\gamma \gamma$ production. However, an explicit numerical calculation shows that the cross sections for $\eta'\eta'$ (and also for $\eta\eta$) CEP after branching to the $4\gamma$ final state are in fact predicted to be a small fraction of the direct $\gamma\gamma$ CEP cross section for the relevant event selection, with a similar result holding for the $\eta\eta'$ final state, see~\cite{HarlandLang:2013} for more details.
 
\begin{table}
\begin{center}
\begin{tabular}{|l|c|c|c|}
\hline
$a_2^G(\mu_0^2)$&-9.5&0&9.5\\
\hline
$\sigma(\eta'\eta')/\sigma(\eta\eta)$&210&1300&1600\\
\hline
$\sigma(\eta'\eta')/\sigma(\gamma\gamma)$&3.5&100&660\\
\hline
$\sigma(\eta\eta)/\sigma(\pi^0\pi^0)$&2.7&12&66\\
\hline
\end{tabular}
\caption{Ratios of $\eta(')\eta(')$ and $\pi^0\pi^0$, $\gamma\gamma$ CEP cross sections at $\sqrt{s}=1.96$ TeV with MSTW08LO PDFs~\cite{Martin:2009iq}, for a $gg$ distribution amplitude with different choices of $a_2^G(\mu_0^2=1\,{\rm GeV}^2)$, corresponding to different normalizations of the $gg$ distribution amplitude $\phi_G(x,Q^2) \propto a_2^G(Q^2)$, guided by the fit of~\cite{Kroll:2012hs}. The $q\overline{q}$ distribution amplitude is given by the CZ form (see~\cite{Benayoun:1989ng}). The meson/photons are required to have transverse energy $E_\perp>2.5$ GeV and pseudorapidity $|\eta|<1$.}\label{etarats1}
\end{center}
\end{table}
 
 \item We recall that as well as having valence $q\overline{q}$ components, it is well known that the dominantly flavour--singlet $\eta'$ (and also, through mixing, $\eta$) mesons should have a valence $gg$ component, which carries flavour--singlet quantum numbers. The CEP of $\eta'$ and $\eta$ meson pairs, in the perturbative regime, represents a novel (and complementary) probe of the size of a $gg$ component of these mesons. In particular, the contribution from a $gg$ valence component of the $\eta'$, $\eta$ mesons enters at the same (leading) order as the purely $q\overline{q}$ contribution in the CEP cross section. An explicit numerical calculation shows that any sizeable $gg$ component of the $\eta'$ (and $\eta$) can have a strong effect on the CEP cross section, increasing (or decreasing) it by up to $\sim$ an order of magnitude, depending on the specific size and sign of the $gg$ component. We show this in Fig.~\ref{etamcz}, taken from~\cite{HarlandLang:2013}, where we plot the $M_X$ distribution for $X=\eta'\eta'$ CEP at $\sqrt{s}=1.96$ TeV for a band of possible $gg$ components, taking the `Chernyak--Zhitnisky' (CZ) form~\cite{Chernyak:1981zz} for the quark distribution amplitude in (\ref{amp}). Although the correct form of the quark distribution amplitude remains an open question, it is shown in~\cite{Chernyak:2009dj} to describe the $\gamma\gamma \to M\overline{M}$ data quite well, and we take it as our benchmark choice here. At the LHC, we expect the cross section (for the same event selection) to be roughly a factor of $\sim 3$--5 larger for $\sqrt{s}=7$--14 TeV, with the particle distributions almost unchanged.
 
  \begin{figure}
\begin{center}
\includegraphics[scale=0.6]{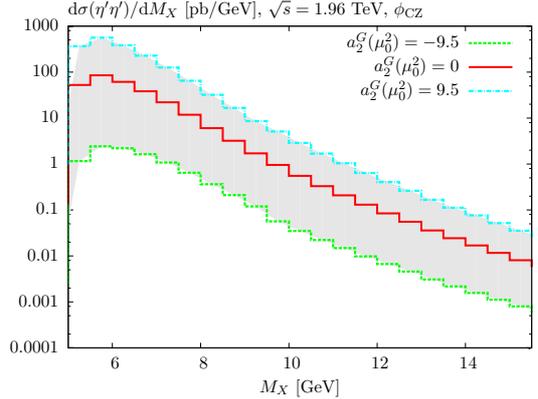}
\caption{Differential cross section ${\rm d}\sigma/{\rm d}M_X$ for $X=\eta'\eta'$ production  at $\sqrt{s}=1.96$ TeV with MSTW08LO PDFs~\cite{Martin:2009iq}, taking the CZ form (see~\cite{Benayoun:1989ng}) for the quark distribution amplitude, and for a band of $a_2^G(\mu_0^2=1\,{\rm GeV}^2)$ values, corresponding to different normalizations of the $gg$ distribution amplitude $\phi_G(x,Q^2) \propto a_2^G(Q^2)$, guided by the fit of~\cite{Kroll:2012hs}. The mesons are required to have transverse energy $E_\perp>2.5$ GeV and pseudorapidity $|\eta|<1$.}\label{etamcz}
\end{center}
\end{figure}
 
 \item Due to the identical angular dependence of the amplitudes (\ref{lad0}), (\ref{tgq0}) and (\ref{tgg0}), the effect of including a non--zero $gg$ component  of the $\eta'$ ($\eta$) mesons on the $\eta(')\eta(')$ CEP amplitudes will to first approximation be to multiply them by an overall normalization factor. The ratio of the different $\eta(')\eta(')$ cross sections are then determined by the mixing parameters (taking the fit of~\cite{Feldmann:1998vh})
\begin{align}\nonumber
\sigma(\eta'\eta'):\sigma(\eta\eta'):\sigma(\eta\eta)&=1:2\tan^2(\theta_1):\tan^4(\theta_1)\;,\\ \label{ratcross}
&\approx 1:\frac{1}{19}:\frac{1}{1450}\;,
\end{align}
 irrespective of the size of the $gg$ component. In fact, the true ratios are expected to deviate somewhat from these values, due to the $|J_z|=2$ contribution from the amplitude (\ref{T+-}), which can in particular be important in the case of the $\eta\eta$ cross section, where the suppression of this contribution, due mainly to the $J_z=0$ selection rule, and that of the flavour--singlet components, due to the small mixing angle $\theta_1$, can be comparable.
 
 \end{enumerate}
 
 Finally, we note that, although it has not been discussed above, theoretical studies of meson pair CEP in fact have a long history, which predate the perturbative Durham approach, see~\cite{HarlandLang:2012qz} and references therein for more details and~\cite{Lebiedowicz:2012nk} for some recent studies.  However, as discussed in~\cite{HarlandLang:2012qz,HarlandLang:2010ys}, at comparatively large meson transverse momenta, $k_\perp$, CEP should be dominated by the perturbative 2--gluon exchange mechanism discussed above and shown in Fig.~\ref{fig:pCp}. At lower $k_\perp$ a study of the transition region between these `non--perturbative' (calculated in a Regge theory framework, see Fig.~5 of~\cite{HarlandLang:2012qz}) and `perturbative' regimes may be necessary. This was performed in~\cite{HarlandLang:2012qz} for the case of $\pi\pi$ CEP. For example, for the ($E_\perp>2.5$ GeV, $|\eta|<1$) meson pair event selection we have considered above, the perturbative contribution is expected to be dominant in all cases.
 
The predicted cross sections are in general much larger in the lower $k_\perp$ ($M_X$) region, for which the perturbative approach is not necessarily applicable. The first data on exclusive $\pi^+\pi^-$ production are already available from CDF, which has produced a measurement of the dipion mass distribution at $M_{\pi\pi}<5.5$ GeV at 0.9 and 1.96 TeV~\cite{cdfdat}. In this case a measurement of the pion $k_\perp$-- distributions\footnote{More specifically, it is the size of the vector difference \linebreak[4]$|({\bf k}_\perp(\pi^+)-{\bf k}_\perp(\pi^-))|/2$ which is important, so that the effect of any non--zero $k_\perp$ of the $\pi\pi$ system due to the $p_\perp$ of the outgoing protons is subtracted.} corresponding to the same data would be very useful, both as a way to probe the (poorly constrained) form factor $F_M(\hat{t})$, for the off--shell intermediate exchange, which enters into the non--perturbative models, as well as the onset of any perturbative (i.e. power--like) behaviour in the higher $k_\perp$ region, which the pQCD--based model discussed above predicts\footnote{A comparison of the $k_\perp$ (and $M_{\pi\pi}$) distributions at 0.9 and 1.96 TeV would probe the size of any possible contamination due to proton dissociation~\cite{HarlandLang:2013jf}. It would also be interesting to measure the distribution in $\Delta y$, the rapidity difference between the pions. This would give information about the spin (or more generally intercept of the Regge trajectory) of the $t$--channel exchange $M^*$.}. We also note that we expect $\pi^+\pi^-$ (and $K^+K^-$\footnote{While the $\chi_{c(0,2)}$ branching ratios to $\pi\pi$ and $KK$ are the same, the exclusive continuum background is in fact expected to be somewhat smaller in the $KK$ case, see~\cite{HarlandLang:2012qz}. Moreover, this mode may be experimentally more favourable, in particular in terms of suppressing the non--exclusive (i.e. with proton dissociation) background, due to the generally higher particle ID efficiency in the Kaon case.}) data to come soon from CMS~\cite{enterria}, with a veto applied on any additional particles within their rapidity coverage, which should contain a sizeable exclusive component.
 
A further motivation for studying the CEP of meson pairs is as a continuum background to exclusive resonant $\chi_c \to \pi\pi$ and $\chi_c \to KK$ production. In the experimentally relevant kinematic region ($M_{\pi\pi}\sim M_\chi$, $k_\perp(\pi)\sim M_\chi/2$), it was found in~\cite{HarlandLang:2012qz} that the non--perturbative contribution discussed above is expected to be dominant, but that with suitable cuts on the meson $k_\perp$ and rapidity $y$, it should represent a realistic channel to observe $\chi_{c0}$ production (both $\chi_{c1}$ and $\chi_{c2}$ production are expected to be suppressed by the $J_z=0$ selection rule). Such studies are, for instance, in the LHCb programme~\cite{lhcb}. We refer the reader to~\cite{HarlandLang:2012qz,HarlandLang:2009qe} for more details.

\section{Conclusion}

The study of the CEP of meson pairs ($M\overline{M}=\pi\pi$, $KK$, $\eta(')\eta(')$...) has provided a large number of interesting and new results, both phenomenological, in terms of their implications for the CEP cross sections, and theoretical, in terms of the many interesting properties of the relevant $gg \to M \overline{M}$ subamplitudes, calculated using a novel application of the `hard exclusive' formalism. In this note we summarize and discuss these results, concentrating on the most important theoretical and phenomenological results and predictions, without entering into the detailed treatment that can be found in~\cite{HarlandLang:2011qd,HarlandLang:2013}. On the experimental side, we emphasize that the CEP of meson pairs is an active field, with existing CDF data on tape and future special low--pile--up runs at the LHC offering the possibility of observing the processes and testing the many non--trivial predictions presented here.

\bibliography{ggbib}
\bibliographystyle{elsarticle-num}

\end{document}